\documentclass[aip,jcp,twocolumn,superscriptaddress,10pt]{revtex4-1}
\usepackage{amsmath}
\usepackage{amsfonts}
\usepackage{mathrsfs}
\usepackage{graphicx,color}
\usepackage{bm,bbm}
\usepackage{multirow}
\usepackage{hyperref}
\usepackage{physics}
\usepackage{enumitem}
\usepackage{cases}
\usepackage[caption=false]{subfig}
\usepackage{algorithm,algpseudocode}

\makeatletter
\newcommand\footnoteref[1]{\protected@xdef\@thefnmark{\ref{#1}}\@footnotemark}
\makeatother

\DeclareMathAlphabet{\mathdj}{U}{msb}{m}{n}

\begin{document}

\title{Solution of Disordered Microphases in the Bethe approximation}
\author{Patrick Charbonneau}
\affiliation{Department of Chemistry, Duke University, Durham, North Carolina 27708, USA}
\affiliation{Department of Physics, Duke University, Durham, North Carolina 27708, USA}
\author{Marco Tarzia}
\affiliation{LPTMC, CNRS-UMR 7600, Sorbonne Universit\'e, 4 Pl. Jussieu, F-75005 Paris, France}
\affiliation{Institut Universitaire de France, 1 rue Descartes, 75231 Paris Cedex 05, France}
\date{\today}

\begin{abstract}
The periodic microphases that self-assemble in systems with competing short-range attractive and long-range repulsive interactions are structurally both rich and elegant. Significant theoretical and computational efforts have thus been dedicated to untangling their properties. By contrast, disordered microphases, which are structurally just as rich but nowhere near as elegant, have not been as carefully considered. Part of the difficulty is that simple mean-field descriptions make a homogeneity assumption that washes away all of their structural features. Here, we study disordered microphases by exactly solving a SALR model on the Bethe lattice. By sidestepping the homogenization assumption, this treatment recapitulates many of the key structural regimes of disordered microphases, including particle and void cluster fluids as well as gelation. This analysis also provides physical insight into the relationship between various structural and thermal observables, between criticality and physical percolation, as well as between glassiness and microphase ordering. 
\end{abstract}
\maketitle

\section{Introduction} 
Periodic microphases (or mesophases) generically form in systems described by competing (effective) short-range attractive and long-range repulsive (SALR) interactions. Similarly ordered structures have thus been reported in materials as diverse as block copolymers, surfactants, colloidal suspensions, cell nuclei, and magnetic alloys.\cite{SA95,GBR12,CPG13,MCMO19,CZ21,riess2003micellization,bates2000block,bates1990block,portmann2003inverse}
The material breadth and interest of this universality class has motivated the development of an extended array of field theoretic, density functional, liquid state, and molecular simulations descriptions (see, \emph{e.g.}, Refs.~[\onlinecite{CZ21},\onlinecite{ZC16,LX19,leibler1980theory}]). Numerical simulations and experiments suggest that disordered microphases are also structurally quite rich, notably exhibiting cluster fluids of both particles and voids as well as equilibrium physical gels.\cite{GR04,IR06,AW07,SPG17,ZC17,ZC21,mani2014equilibrium,godfrin2014generalized,stradner2004equilibrium,campbell2005dynamical,de2006columnar,toledano2009colloidal,sciortino2005one,sciortino2004equilibrium}  Disordered microphases-based materials have even found technological applications as filtration membranes.\cite{YQMMC15} 
As equilibrium precursors to periodic microphase formation, they are also of clear self-assembly interest.

From the theoretical standpoint, disordered microphases have received much less attention than their ordered (periodic) counterparts. Standard mean-field descriptions simply wash away all structural features of the high-temperature phase. Because disordered microphases are spatially homogeneous on average their instantaneous density inhomogeneities do not naturally emerge from standard treatments. Density-functional descriptions, for instance, would require the consideration of higher-order density terms beyond what current approaches do\cite{Ar08,CPG13} in order to carve out the relevant structural properties.  In liquid-state descriptions as well, the disordered microphase regime is not fully captured.\cite{AW07} As a result the various structural regimes of disordered microphases are mostly described phenomenologically, especially for particle cluster (or micelle) formation.\cite{LOW83,Wi04,BH03,GBR12,HG20}

A different road toward exact solutions of SALR models proceeds through the consideration of Cayley trees with finite connectivity $c+1>2$, for which the Bethe approximation is exact. While the mean-field nature of these models limits the extent to which they recapitulate finite-dimensional physics, their finite yet non-trivial connectivity preserves a notion of distance between lattice sites as well as local short-range correlations that are key to disordered microphases. They therefore naturally bypass the homogenization assumption of mean-field treatments. Surprisingly, while the low-temperature periodic microphases of a variety of such models were studied over a generation ago,\cite{Va81,IT83,YOS85,MCA85,SC86,RAU14} their disordered microphases were not similarly considered, possibly because the theoretical machinery then available was not yet fully developed.\cite{MP01,MM09} We here remediate this oversight to gain material insight into disordered microphases. More specifically, we exactly solve a model SALR Hamiltonian on locally tree-like graphs (Cayley-tree--like) to investigate the clustering crossover and its connection with the peak of the heat capacity, as well as the interplay between percolation, glass formation and the ordered microphase regime.

The plan for rest of this article is as follows. Section~\ref{sec:model} describes the specific SALR model considered, Sec.~\ref{sec:cavitysoln} introduces the cavity field equations to solve this model in the Bethe approximation, and Sec.~\ref{sec:homo} and \ref{sec:inhomogeneous} describe the various schemes used to study homogeneous and inhomogeneous phases, respectively. Results are discussed in Sec.~\ref{sec:discussion}, and a brief conclusion follows in Sec.~\ref{sec:conclusion}.

\section{Model}
\label{sec:model}
\begin{figure*}
	\includegraphics[width=1.1\columnwidth]{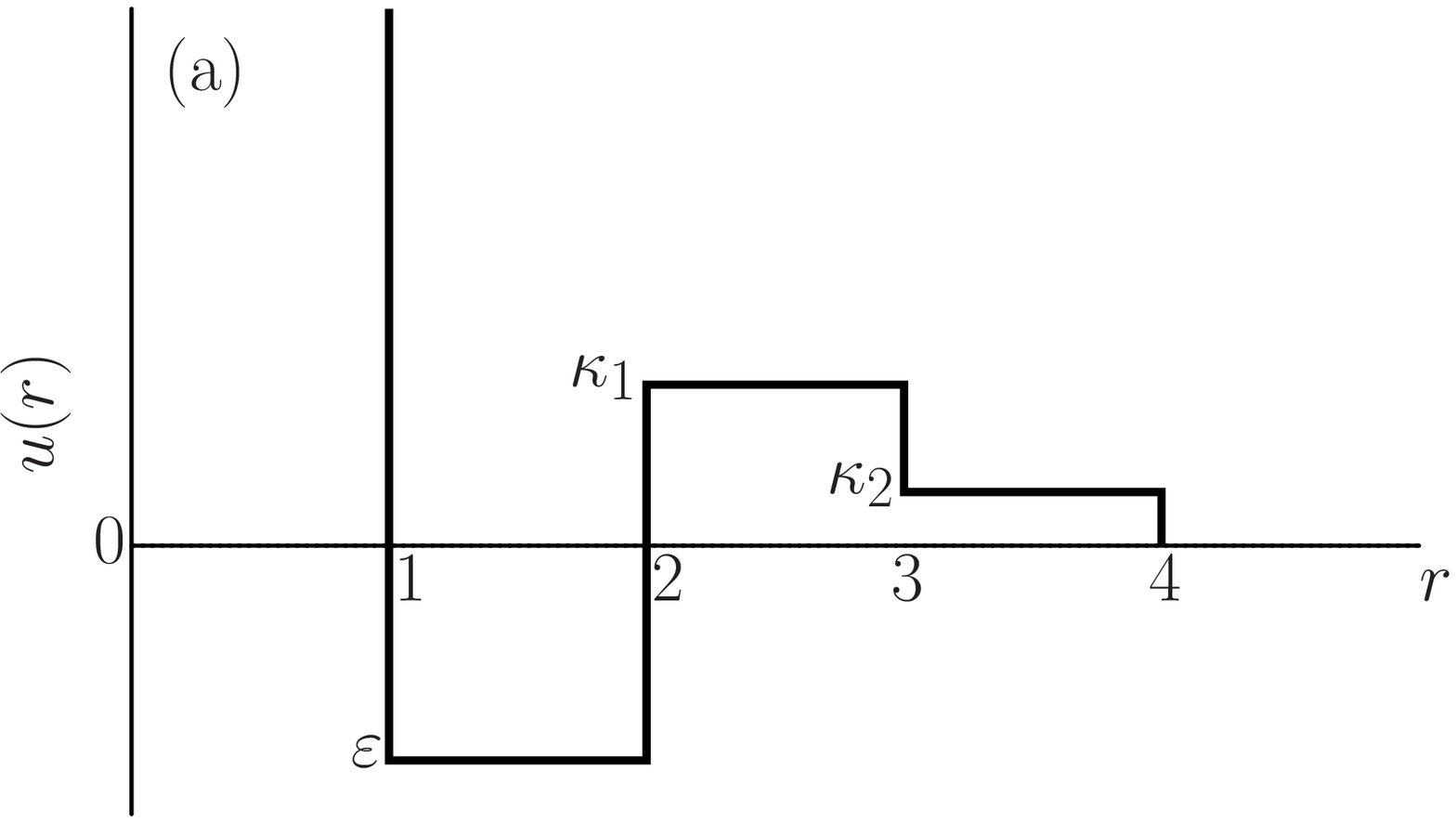}
	\includegraphics[width=0.9\columnwidth]{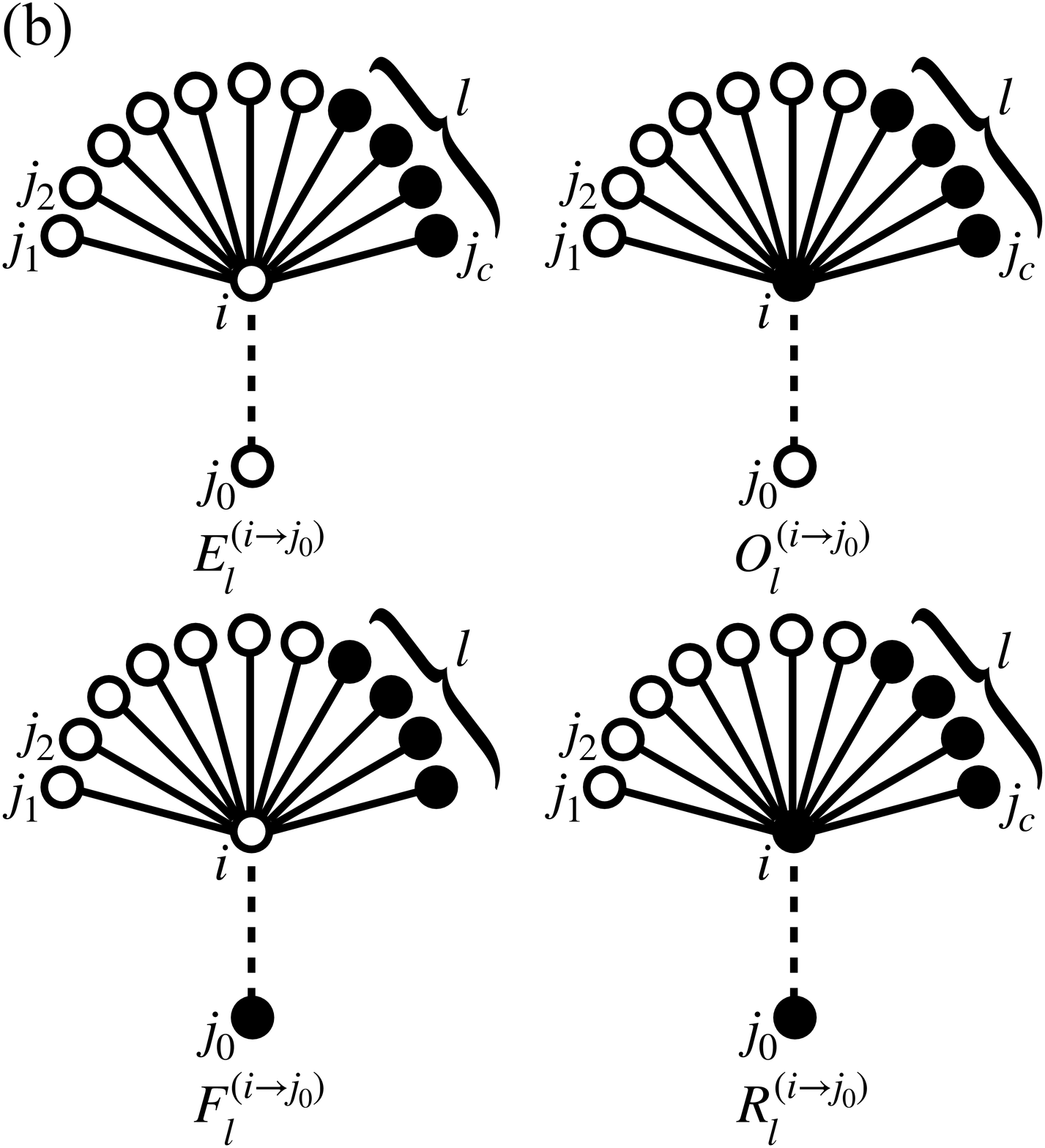}
	\caption{(a) Schematic SALR radial interaction potential given by Eq.~\eqref{eq:H} as a function of Hamming distance $r$ on the graph. In continuous space, the lattice SALR interaction considered here is akin to a square-well attraction of depth $\varepsilon$ with two repulsive steps of strength $\kappa_1$ and $\kappa_2$ (not drawn to scale), respectively. (b) Illustration of the configurations corresponding to the cavity fields defined in Eqs.~\eqref{eq:messagepassing} for $c=11$ and $l=4$.}
    \label{fig:cavity_fields}
\end{figure*}
The model Hamiltonian is expressed in terms of occupation variables,  $n_i = 0,1$,
\begin{equation} \label{eq:H}
{\cal H} = - \varepsilon \sum_{\langle i, j \rangle} n_i n_j + \kappa_1 \! \! \sum_{\langle \langle i, j \rangle \rangle} \!\! n_i n_j  + \kappa_2 \! \! \! \sum_{\langle \langle \langle i, j \rangle \rangle \rangle } \! \!\! n_i n_j  - \mu \sum_i n_i\, ,
\end{equation}
where the first term encodes the nearest-neighbor attraction with $\varepsilon>0$ setting the unit of energy, and the second and third terms encode next- and next-next-nearest-neighbor repulsion, respectively (see Fig.~\ref{fig:cavity_fields}).\cite{SC86} The last term is akin to a uniform external field, which in this representation is equivalent to fixing the chemical potential, and hence to tuning the system density. We consider this model on a (locally) tree-like lattice of fixed connectivity $c+1$ (see below for a precise definition). In order to remain in a regime akin to what is observed in three-dimensional simulations,\cite{ZC21} we thus set
\begin{equation} \label{eq:potential}
\left \{
\begin{array}{l}
\kappa_1 = \kappa\varepsilon \, ,\\
\kappa_2 =  \kappa \varepsilon / (c+1) \, ,
\end{array}
\right .
\end{equation}
with $\kappa$ parameterizing the overall repulsion strength, and the $c+1$ factor compensating for the exponential (instead of algebraic) growth of the number of neighbors with distance on tree-like (instead of real-space) lattices.

Note that by going from occupation to spin variables, i.e., $s_i = 2 n_i - 1$, it is straightforward to show that the $Z_2$ symmetry of the model is restored for 
\begin{equation} \label{eq:mu05}
\mu_0 = (c+1)(- \varepsilon + c \kappa_1 + c^2 \kappa_2)/2    
\end{equation}
This choice is indeed equivalent to canceling the external magnetic field for spin variables, and thus ensures that the average density $\rho = \langle n _i \rangle = 1/2$. This symmetry reveals that the independent density range is $\rho\in[0,1/2]$. For the $Z_2$ symmetric case it is also possible to calculate the frustration $\kappa_0$ at which the energetic ground state transitions from gas-liquid coexistence (ferromagnetic) to layered with a three-fold periodicity ($\langle 3\rangle$for layers with a period $\lambda_\ell=6$). For $\mu$ given by Eq.~\eqref{eq:mu05}, the energy per site of such lamellar ground state is indeed
\begin{equation}
e_\ell = \frac{\varepsilon}{6} - \frac{\kappa_1 c}{3} - \frac{\kappa_2}{6} (3 c^2 - 2 c + 2) \, .
\end{equation}
The point at which $e_\ell$ equates the gas-liquid coexistence ground state energy per site---$e_{f}=0$ for Eq.~\eqref{eq:potential}---is then
\begin{equation}
\label{eq:T0mod}
\kappa_0 = \frac{c + 1}{5c^2 + 2} \ .
\end{equation}

Recall that the Bethe approximation (see Sec.~\ref{sec:cavitysoln}) was originally introduced for Cayley trees of fixed connectivity $c+1$, which are loop-less graphs with a finite fraction of sites lying on the boundary, \emph{i.e.}, tree leaves. This hierarchical structure leads to exact recursion relations that can be solved iteratively for a given boundary condition on these leaves. When the Gibbs measure is characterized by more than one minimum, however, the fixed point of this recursion can be strongly affected by the choice of boundary condition, due to the significant contribution of leaves, even in the thermodynamic limit of $N\to\infty$ nodes.\cite{MP01} One way to sidestep the issue is to define the lattice as a random-regular graph (RRG) of fixed connectivity $c+ 1$ with no trivial loops (joining a site to itself) nor multi-edges (distinct edges joining the same sites). Extensive studies of such graphs have indeed revealed that in the limit $N\to\infty$, typical RRGs have loop lengths $\mathcal{O}(\ln N)$.\cite{RRG} They are therefore locally tree-like, which makes the Bethe approximation (locally) asymptotically exact. In addition, the large loops implement an analog of 
self-consistent boundary conditions without having to rely on external constraints. The resulting frustration, however, forbids the formation of the long-range periodically modulated structures expected for  SALR model on Cayley trees at large $\kappa$ and low $T$.\cite{Va81,IT83,YOS85,MCA85,SC86,RAU14} This last aspect is further explored in Sec.~\ref{sec:inhomogeneous}.

\section{Cavity fields and recursion relations}
\label{sec:cavitysoln}

The cavity method is the standard approach for solving a model such as Eq.~\eqref{eq:H} on a RRG. This recursive scheme entails first selecting a node $i$ of the lattice, for which one of the $c+1$ edges---say the edge it has with node $j$, $i \leftrightarrow j$---is removed, and hence the cavity site $i$ roots a semi-infinite branch of the tree. (For convenience and without loss of generality, we denote the missing edge $i \leftrightarrow j$ as the \emph{backward} edge.) Taking $c$ cavity sites and connecting them to a new site through $c$ edges produces a new cavity site with the  same statistical properties as its $c$ neighbors. This procedure thus gives rise to exact (in the thermodynamic limit) recursion relations for the cavity fields. These are defined as the local marginal probabilities of having specific configurations of the occupation variables on a given cavity site, once all other degrees of freedom on the branch have been integrated out.
This feat is possible thanks to the tree-like structure of RRGs, which implies that the $c$ neighbors of a given sites $i$ are uncorrelated in the absence of site $i$, and thus have factorizable joint probabilities. (See Ref.~[\onlinecite{MM09}] for a detailed presentation of all facets and subtleties of the cavity method in the context of optimization problems and information theory, and Sec.~IIIA of Ref.~[\onlinecite{STZ09}] for a pedagogical explanation of the method for the ferromagnetic Ising model.)

We now specialize to the model given by Eq.~\eqref{eq:H}. For notational convenience, we introduce the variable $w_{i \to j} = \sum_{m \in \partial i / j} n_m$, which counts the number of occupied neighbors of the cavity site $i$ in absence of its backward neighbor $j$.
The cavity fields are then defined as the probabilities of having different occupancy configurations of the cavity sites and of their neighbors (see Fig.~\ref{fig:cavity_fields} for an illustration, and Ref.~[\onlinecite{CFT20}] for a similar calculation for a model with next-next-nearest neighbor interactions): 
\begin{align}
\left\{
\begin{array}{l}
\label{eq:messagepassing}
E_l^{(i \to j)} \equiv \textrm{Prob}(n_i=0 \textrm{ 
\& } n_j=0 \textrm{ \& } w_{i \to j} = l) \\
F_l^{(i \to j)} \equiv \textrm{Prob}(n_i=0 \textrm{ 
\& }n_j=1  \textrm{ \& } w_{i \to j} = l) \\
O_l^{(i \to j)} \equiv \textrm{Prob}(n_i=1 \textrm{ 
\& }n_j=0  \textrm{ \& } w_{i \to j} = l) \\
R_l^{(i \to j)} \equiv \textrm{Prob}(n_i=1 \textrm{ 
\& }n_j=1  \textrm{ \& } w_{i \to j} = l)
\end{array}
\right .
\end{align}
For example, $E_l^{(i \to j)}$ is the probability that cavity site $i$ is empty ($n_i=0$), with its backward site $j$ also empty ($n_j=0$) and with $l$ occupied neighbors  (with $0 \le l \le c$), as depicted in Fig.~\ref{fig:cavity_fields} for $c=11$ and $l=4$.

In order to write compact recursion relations for these objects, we also introduce auxiliary functions of the local marginal probabilities and of the variable $q$ at inverse temperature $\beta=1/T$ (with Boltzmann constant $k_B=1$)
\begin{equation} \label{eq:aux}
\begin{aligned}
\hat{\varphi}_q^{(i \to j)} &= \sum_{m=0}^c \varphi_m^{(i \to j)} e^{- m q \beta \kappa_2} \, , 
\end{aligned}
\end{equation}
where $\varphi_m = \{ E_m,F_m,O_m,R_m \}$ corresponds to the different kinds of cavity fields defined in Eqs.~\eqref{eq:messagepassing}.
Considering the iteration process in which $c$ cavity sites $\{j_1, \ldots , j_c \}$ are connected to a new  cavity site $i$ through $c$ edges, we have:
\begin{widetext}
\begin{equation} \label{eq:cavity_inhomogeneous}
\begin{aligned}
E_l^{(i \to j_0)}  =& \left ( Z_{\rm iter}^{(i \to j_0)} \right)^{\!-1} e^{- \beta \frac{l (l-1)}{2} \kappa_1} \!\!\!\sum_{1 \le j_1 < \ldots < j_l \le c} \left[  \prod_{k=1}^l \hat{O}_{l-1}^{(j_k \to i)} \!\!\!\prod_{j_q \notin \{ j_1 ,  \ldots , j_l \}} \!\!\! \hat{E}_l^{(j_q \to i)} \right]  \, , \\
F_l^{(i \to j_0)}  =& \left ( Z_{\rm iter}^{(i \to j_0)} \right)^{\!-1}  e^{- \beta \frac{l (l+1)}{2}  \kappa_1} \!\!\!\sum_{1 \le j_1  < \ldots < j_l \le c} \left[ \prod_{k=1}^l \hat{O}_l^{(j_k \to i)}  \!\!\!\prod_{j_q \notin \{ j_1 ,  \ldots , j_l \}}\!\!\! \hat{E}_{l+1}^{(j_q \to i)} \right] \, , \\
O_l^{(i \to j_0)}  =&  \left (Z_{\rm iter}^{(i \to j_0)} \right)^{\!-1} e^{\beta \left[ \mu + l \left( \varepsilon  - \frac{(l-1)}{2} \kappa_1 \right) \right]} \!\!\!\sum_{1 \le j_1 < \ldots < j_l \le c} \left[ \prod_{k=1}^l \hat{R}_{l-1}^{(j_k \to i)} \!\!\!\prod_{j_q \notin \{ j_1 ,  \ldots , j_l \}} \!\!\! \hat{F}_l^{(j_q \to i)} \right] \, , \\
R_l^{(i \to j_0)}  =& \left (Z_{\rm iter}^{(i \to j_0)} \right)^{\!-1} e^{\beta \left[ \mu + l \left( \varepsilon  - \frac{(l+1)}{2} \kappa_1 \right) \right]} \!\!\!\sum_{1 \le j_1 < \ldots < j_l \le c} \left[ \prod_{k=1}^l \hat{R}_{l}^{(j_k \to i)} \!\!\!\prod_{j_q \notin \{ j_1 ,  \ldots , j_l \}} \!\!\! \hat{F}_{l+1}^{(j_q \to i)} \right] \, ,
\end{aligned}
\end{equation}
where the normalization factor $Z_{\rm iter}^{(i \to j_0)}$ is such that $\sum_{l=0}^c (E_l^{(i \to j_0)} + F_l^{(i \to j_0)} + O_l^{(i \to j_0)} + R_l^{(i \to j_0)}) = 1$.
As discussed above, these equations become asymptotically exact in the thermodynamic limit, $N\to\infty$, on random sparse lattices with a local tree-like structure (such as RRGs) and they are exact for loop-less Cayley trees even at finite $N$. For a generic tree-like graph of $N$ nodes and connectivity $c+1$, Eqs.~\eqref{eq:cavity_inhomogeneous} correspond to a system of $4 (c+1)^2 N$ coupled non-linear equations for the $4 (c+1)$ cavity fields defined on all the $N (c+1)$ (directed) edges of the graph.

These local probabilities, however, are defined on intermediate objects (the cavity sites) with one fewer link and one fewer neighbor than nodes of the original lattice, and hence are statistically different from them. Most of the thermodynamic observables of the original Bethe lattice, such as the free energy, the average density, the average energy, the specific heat, \emph{etc}., can nonetheless be computed from the fixed point of the self-consistent Eqs.~\eqref{eq:cavity_inhomogeneous}.\cite{MM09,STZ09} In order to do so, one considers the process by which $c+1$ cavity sites $\{j_1, \ldots , j_{c+1} \}$ are connected to a central site $i$. Defining ${\cal E}_i$ and ${\cal O}_i$ as the probabilities that a given site is empty or occupied, respectively, we obtain
\begin{equation} \label{eq:final_inhomogenous}
\begin{aligned}
{\cal E}_i & = \left ( Z_{\rm site}^{(i)} \right)^{\!-1} \sum_{l=0}^{c+1} e^{- \beta \frac{ l (l-1)}{2} \kappa_1} \left \{ \sum_{1 \le j_1 <  \ldots < j_l \le c+1} \left[ \prod_{k=1}^l \hat{O}_{l-1}^{(j_k \to i)}  
\prod_{j_q \notin \{ j_1 ,  \ldots , j_l \}} \!\!\! \hat{E}_l^{(j_q \to i)} \right ] \right\}\, , \\
{\cal O}_i & =  \left ( Z_{\rm site}^{(i)} \right)^{\!-1}  \sum_{l=0}^{c+1} e^{\beta \left[ \mu +l\left( \varepsilon - \frac{l-1}{2} \kappa_1 \right) \right]} \left \{ \sum_{1 \le j_1 <  \ldots < j_l \le c+1} \left[  \prod_{k=1}^l \hat{R}_{l-1}^{(j_k \to i)} \!\!\!\prod_{j_q \notin \{ j_1 ,  \ldots , j_l \}} \!\!\! \hat{F}_l^{(j_q \to i)} \right ] \right \} \, , 
\end{aligned}
\end{equation}
where the normalization factor $Z_{\rm site}^{(i)}$ is such that ${\cal E}_i + {\cal O}_i = 1$.
The average density on site $i$ is then $\rho_i = {\cal O}_i/({\cal E}_i + {\cal O}_i) = {\cal O}_i$.
In order to compute the system free energy, we follow Refs.~[\onlinecite{MM09},\onlinecite{STZ09},\onlinecite{RBMM04}] in applying the following construction:
\begin{itemize}
\item Start with a $c+1$--connected Bethe lattice of $N$ nodes;
\item For each node, pick and remove an edge $i \leftrightarrow j$, leading to
$2(c + 1)$ cavity sites; 
\item Form two Bethe lattices of $N+1$ nodes by adding two new sites and connecting each to $c + 1$ cavity sites. 
\end{itemize}
Because two sites were added, the free energy difference between the resulting lattice with  $N+1$ nodes and the initial one with $N$ nodes is then simply twice the free energy per site. Averaging over all possible choices of removed edges, the free energy per site can then be written as
\begin{equation} \label{eq:free}
f = \frac{1}{N}\left(\sum_i \Delta F_{\rm site}^{(i)} - \sum_{\langle i, j \rangle} \Delta F_{\rm link}^{(i \leftrightarrow j)}\right) \, 
\end{equation}
where $e^{-\beta \Delta F_{\rm site}^{(i)}}  = Z_{\rm site}^{(i)}$ and 
\begin{equation}
\begin{aligned}
 e^{-\beta \Delta F_{\rm link}^{(i \leftrightarrow j)}} & = \sum_{m_1,m_2=0}^c \left(E_{m_1}^{(i \to j)} E_{m_2}^{(j \to i)} + F_{m_1}^{(i \to j)} O_{m_2}^{(j \to i)} + F_{m_1}^{(j \to i)} O_{m_2}^{(i \to j)} + R_{m_1}^{(i \to j)} R_{m_2}^{(j \to i)} \right ) e^{- m_1 m_2 \beta \kappa_2} \, ,
\end{aligned}
\end{equation}
are the free energy shifts due to the addition of site $i$ and of edge $i \leftrightarrow j$, respectively.
\end{widetext}
Three kinds of relevant physical solutions of the recursion Eqs.~\eqref{eq:cavity_inhomogeneous} and~\eqref{eq:final_inhomogenous} are possible:
\begin{itemize}
    \item Translationally invariant solutions in which the cavity fields do not fluctuate from site to site. These homogeneous solutions are described in Sec.~\ref{sec:homo} and correspond to the equilibrium disordered microphases (paramagnetic phase), which are the main focus of this work, as well as to the gas-liquid coexistence region (ferromagnetic phase);
    \item Layered or periodic solutions in which different generations carry different cavity fields in a modulated way. These solutions break translational invariance and correspond to the periodic microphases described in Sec.~\ref{sec:inhomogeneous}.\cite{Va81,IT83,YOS85,MCA85,SC86,RAU14}
    \item Disordered glassy solutions in which the cavity fields fluctuate from one site to another without periodicity. Such glassy phases have been found at low temperatures and high packing fractions in lattice glass models on locally tree-like graphs with short-ranged hard-core interactions.\cite{BM02,CTCC03,CTCC03b,T07,RBMM04,KTZ08} A low temperature glass transition has also been reported in SALR models\cite{SW00,HSW01,GTV02,GTV02b,TC06,TC07} in the framework of a Ginzburg-Landau description and a mean-field approximation, but has not been evaluated in the context of lattice models. In the glassy regime the cavity fields are random variables and Eqs.~\eqref{eq:cavity_inhomogeneous} must be interpreted as recursive relations for their probability distributions. In this article, we demonstrate the existence of a glassy phase in some region of the phase diagram of the family of SALR models described by Eq.~\eqref{eq:H}, but leave this regime for further study.
\end{itemize}
\section{Homogeneous solutions}
\label{sec:homo}
The disordered microphases, which are the main focus of this work, correspond to the paramagnetic phase and hence are translationally invariant. Cavity fields then do not depend on the chosen cavity site $i$, and thus give rise to homogeneous solutions of Eq.~\eqref{eq:cavity_inhomogeneous}. In this case, the recursion relations simplify to:
\begin{equation} \label{eq:cavity}
\begin{aligned}
E_l & = Z_{\rm iter}^{-1} e^{- \beta \frac{l (l-1)}{2} \kappa_1} {{c}\choose{l}} 
 \hat{O}_{l-1}^l \hat{E}_l^{c - l}  \, , \\
F_l & = Z_{\rm iter}^{-1} e^{- \beta \frac{l (l+1)}{2} \kappa_1} {{c}\choose{l}} \hat{O}_l^l \hat{E}_{l+1}^{c - l}  \, , \\
O_l & = Z_{\rm iter}^{-1} e^{\beta \left[ \mu + l \left ( \varepsilon - \frac{(l-1)}{2} \kappa_1 \right) \right ] } {{c}\choose{l}} 
\hat{R}_{l-1}^l \hat{F}_l^{c - l}  \, , \\
R_l & = Z_{\rm iter}^{-1}  e^{\beta \left[ \mu +  l \left ( \varepsilon - \frac{(l+1)}{2} \kappa_1 \right ) \right ] }  {{c}\choose{l}}
\hat{R}_l^l \hat{F}_{l+1}^{c - l}  \, ,
\end{aligned}
\end{equation}
where the normalization factor $Z_{\rm iter}$ is such that $\sum_{l=0}^c (E_l + F_l + O_l + R_l) = 1$ and the auxiliary functions $\hat{E}_q,\hat{O}_q,\hat{F}_q,\hat{R}_q$ are defined in Eq.~\eqref{eq:aux}. Equations~\eqref{eq:cavity} then give rise to a system of $4(c+1)$ coupled algebraic equations, whose fixed point of can be straightforwardly determined by iterating numerically. Connecting $c+1$ cavity sites  to a central site gives 
\begin{equation} \label{eq:final}
\begin{aligned}
{\cal E} & = Z_{\rm site}^{-1} \sum_{l=0}^{c+1} e^{- \beta \frac{l (l-1)}{2} \kappa_1} {{c+1}\choose{l}}  \hat{O}_{l-1}^l \hat{E}_l^{c + 1 - l}  \, , \\
{\cal O} & = Z_{\rm site}^{-1} \sum_{l=0}^{c+1} e^{\beta \left[ \mu +  l \left ( \varepsilon - \frac{(l-1)}{2} \kappa_1 \right) \right ] }  {{c+1}\choose{l}} \hat{R}_{l-1}^l\hat{F}_l^{c +1 - l}  \, , 
\end{aligned}
\end{equation}
where the normalization factor $Z_{\rm site}$ is such that ${\cal E} + {\cal O} = 1$.
The free energy per site in Eq.~\eqref{eq:free} then becomes \begin{equation} \label{eq:freegraph}
f = \Delta F_{\rm site} - \frac{c+1}{2} \Delta F_{\rm link} = \Delta F_{\rm iter} - \frac{c - 1}{2} \Delta F_{\rm link} \, ,
\end{equation}
where the second equality stems from site insertion involving the addition of a new cavity site and a link between this cavity site and another cavity site, \emph{i.e.}, $\Delta F_{\rm site} = \Delta F_{\rm iter} + \Delta F_{\rm link}$.
Most of the thermodynamic observables of interest can be computed directly as derivative of the free energy, \emph{e.g.},
\begin{equation}
    \begin{aligned}
     \rho &= - \frac{\partial ( \beta f)}{\partial (\beta \mu)} = \frac{{\cal O}}{{\cal E} + {\cal O}} = {\cal O} \, , \\
     e - \mu \rho & = \frac{\partial (\beta f)}{\partial \beta} \, , \\
     f & = e - T s - \mu \rho 
     \, ,
    \end{aligned}
\end{equation}
where $e=\langle H \rangle/N$ is the average energy per site and $s=S/N$ is the entropy per site.  

\subsection{Percolation}
An important structural feature of the high-temperature phase is the percolation regime, which in microphase formers often results in the formation of a gel-like structure. In order to delineate the boundaries of this regime, we compute the probability that a particle belongs to the percolating cluster ${\cal C}_\infty$. For notational convenience, we introduce the auxiliary probabilities $p_l$ and $q_l$ that the cavity site $i$ (in absence of the backward site $k$) has $w_{i \to k} = l$  occupied neighbors (with $1 \le l \le c$) with the backward site $k$ occupied, and belongs to the semi-infinite percolating cluster on the branch originating from $i$ (see Ref.~[\onlinecite{KTZ08}] for a similar calculation):
\[
\left\{
\begin{array}{l}
p_l \equiv \textrm{Prob} (n_i=1\, \&\, n_k=1\, \&\, 
 w_{i \to k} = l \, \&\,
i\in {\cal C}_\infty) \\
q_l \equiv \textrm{Prob}(n_i=1\, \&\, n_k=1\, \&\, w_{i \to k} = l \, \&\, i\not\in {\cal C}_\infty)
\end{array}
\right .
\]
In order to express these probabilities in terms of cavity fields, we introduce two auxiliary functions
\[
\begin{aligned}
 \hat{p}_q& = p_B \sum_{m=1}^c p_m e^{- m q \beta \kappa_2} \, , \\
 \hat{q}_q& = R_0 + \sum_{m=1}^c (R_m - p_B p_m) e^{- m q \beta \kappa_2} \, . \\
 \end{aligned}
 \]
The probability that two occupied nearest-neighbor sites are bonded, $p_B$, accounts for the percolation of a \emph{physical} cluster, for which one considers a site-bond percolation problem on top of the \emph{geometrical} cluster formed by occupied sites.\cite{CK80} Geometrical clusters  include all nearest-neighbor particles, and are recovered by setting $p_B = 1$; physical clusters control the elastic response of the fluid.\cite{ZZBRDX,XMG04,DdC02} In fact in a simple fluid, the percolation of geometrical clusters, which develops well above the liquid-gas critical temperature, plays no physical role. By contrast, for a suitable choice of $p_B$ percolation of the latter is related to the percolation of critical droplets associated with the gas-liquid phase transition. For the standard lattice gas model ($\kappa = 0$), it has been proven that for $p_B = 1-e^{-\beta \varepsilon/2}$ the percolation line intersect the liquid gas critical point\cite{CK80} (see Fig.~\ref{fig:TcTl} and Fig.~\ref{fig:PD}a). By analogy with the standard lattice gas, we here choose $p_B$ henceforth. The self-consistent equations for $p_l$ and $q_l$ then read
\begin{equation}
\label{eq:pq}
\begin{aligned}
p_l & = Z_p^{-1} e^{\beta \left [ \mu +  l \left ( \varepsilon - \frac{(l+1)}{2} \kappa_1 \right) \right ] } {{c}\choose{l}} 
\hat{F}_{l+1}^{c - l}
\sum_{k=1}^l {{l}\choose{k}} 
\hat{p}_l^k 
\hat{q}_l^{l-k}
\, , \\
q_l & = Z_p^{-1} e^{\beta \left[ \mu +  l \left ( \varepsilon - \frac{(l+1)}{2} \kappa_1 \right) \right ] } {{c}\choose{l}} 
\hat{q}_l^l \hat{F}_{l+1}^{c - l}, 
\end{aligned}
\end{equation}
where the normalization factor $Z_p$ is such that $p_n + q_n = R_n$ and the auxiliary function $\hat{F}_q$ is defined in Eq.~\eqref{eq:aux}. The physical meaning of Eqs.~\eqref{eq:pq} is transparent. In order for a given cavity site to belong to the percolating cluster living on the semi-infinite branch of the lattice originating from it, at least one of its nearest-neighbors must be occupied and belong to the percolating cluster. 

 From the fixed point of Eqs.~\eqref{eq:pq} one can then compute ${\cal P}$ and ${\cal Q}$, defined as the probabilities that a site of the original lattice (with connectivity $c+1$) is occupied and belongs or not, respectively, to the percolating clusters. Connecting $c+ 1$ cavity sites to a central site one finds that these objects satisfy
\begin{equation}
\begin{aligned}
{\cal P} & = {\cal Z}_p^{-1} 
\sum_{l=1}^{c + 1} 
\sum_{k = 0}^{c + 1 - l} e^{\beta \left[ \mu + (l+k) \left ( \varepsilon - \frac{l+k-1}{2} \kappa_1 \right) \right]} \\
 & \,\,\, \times  \frac{(c+1)!}{l! k! (c - l - k + 1)!} 
 \hat{p}_{l+k-1}^l 
\hat{q}_{l+k-1}^k  \hat{F}_{l+k}^{c - l - k + 1} 
\, , \\
{\cal Q} & =  {\cal Z}_p^{-1}\sum_{l=0}^{c+1} 
e^{\beta \left[ \mu + l \left ( \varepsilon - \frac{l-1}{2} \kappa_1 \right) \right]}
{{c+1}\choose{l}} 
\hat{q}_{l-1}^l 
\hat{F}_l^{c + 1 - l} 
\, ,
\end{aligned}
\end{equation}
where the normalization factor ${\cal Z}_p$ is such that ${\cal P} + {\cal Q} = \rho$.
Within the high-temperature phase a transition from a non-percolating phase (${\cal P}=0$) to a percolating phase (${\cal P}>0$) takes place upon decreasing temperature or increasing density (see Fig.~\ref{fig:PD}). In the vicinity of the percolation threshold, ${\cal P}$ behaves critically, scaling as expected for the simple percolation mean-field universality class ${\cal P} \propto \rho - \rho_{\rm perc}$.\cite{SA94}

\subsection{Cluster size distribution} \label{sec:clusters}
In order to identify and study cluster formation, we seek the whole probability distribution $\Pi(s)$ of clusters of $s$ particles. We first illustrate the computation of $\Pi(s)$ for the simple case of site percolation. In this case, each node of the lattice is occupied with probability $p$ and empty with probability $1-p$, which corresponds to the percolation of geometric clusters ($p_B=1$) at infinite temperature ($\varepsilon = \kappa_1 = \kappa_2 = 0$) of our model (with $p = \rho=e^{\beta \mu}/(1 + e^{\beta \mu})$).
For a given cavity node $i$ (once the edge with its backward site has been removed), we define $\pi(s)$ as the probability that, if occupied, $i$ belongs to a cluster of size $s$  on the semi-infinite branch of the lattice originating from it. It is obvious that $s=1$ if and only if all the $c$ neighbors of $i$ are empty, \emph{i.e.}, $\pi(1)=(1-p)^c$. Similarly, for $s=2$ all but one neighbor of $i$ should be empty, and the only occupied node should belong to a cluster of size $1$ once the edge which connects it to $i$ is removed, \emph{i.e.}, $\pi(2)= c (1-p)^{c-1} p \pi(1)$. For $s=3$, two possibilities exist: either $i$ has only one occupied neighbor and this neighbor belongs to a cluster of size $2$ in absence of its edge with $i$, or $i$ has two occupied neighbors with $s=1$, \emph{i.e.},
$\pi(3) = c (1-p)^{c-1} p \pi(2) + c (c-1) (1-p)^{c-2} [p (\pi(1)]^2/2$.
Following this line of reasoning one can easily compute $\pi(s)$ for general $s$ as:
\[
\begin{aligned}
\pi(s) & = \sum_{l=1}^{l_{\rm max}} {{c}\choose{l}} (1-p)^{c-l} \\
& \qquad \qquad \times \prod_{\alpha=1}^{l} \left[ \sum_{s_\alpha} p \pi(s_\alpha) \right] \delta \left (1 + \sum_{\alpha=1}^{l} s_\alpha - s \right ) \, ,
\end{aligned}
\]
where $l_{\rm max}$ is $c$ if $s>c$ and $s-1$ otherwise.
This exercise shows how one can explicitly compute $\pi(s)$ as a function of all the $\pi(s^\prime)$ with $s^\prime<s$ starting from $s=1$ and up to a chosen threshold $s_{\rm max}$. Once these objects are known, the probability $\Pi(s)$ that a given node of the original lattice (in which all the sites have $c+1$ neighbors) belongs to a cluster of size $s \ge 1$ is given by
\[
\begin{aligned}
\Pi(s) & = p \sum_{l=1}^{l_{\rm max}} {{c+1}\choose{l}} (1-p)^{c+1-l} \\
& \qquad \qquad \times \prod_{\alpha=1}^{l} \left[ \sum_{s_\alpha} p \pi(s_\alpha) \right] \delta \left (1 + \sum_{\alpha=1}^{l} s_\alpha - s \right ) \, ,
\end{aligned}
\]
where $l_{\rm max}$ is $c+1$ if $s>c+1$ and $s-1$ otherwise.

The calculation of $\Pi(s)$ for the SALR model described by~Eq.~\eqref{eq:H} is slightly more involved but follows the same logic. In order to take into account the interaction terms of the Hamiltonian one needs to introduce the auxiliary probabilities $\pi_{l,l_1} (s)$ on the cavity site $i$ that the node is occupied, with its backward site occupied, and with $l$ total occupied neighbors, $l_1$ of them belonging to the same clusters ${\cal C}_s$ 
of size $s$ on the semi-infinite branch of the Bethe lattice originating from it. 
The variables $l$, $l_1$, and $s$ must further satisfy the following constraints: 
\[
\begin{aligned}
\theta(s-1) \le l_1 \le {\rm max} (c, s-1) \, , &\\ 
 l_1 \le l \le c \, , &
\end{aligned}
\]
where $\theta(s-1) = 0$ if $s=0$ and $1$ if $s>0$.
Obviously the total number of occupied neighbors must be larger than $l_1$. If $s>1$ then at least one neighbor must belong to ${\cal C}_s$ and thus must be occupied, hence $l_1 \geq 1$. 
The maximum number of neighbors in the same cluster $l_1$ is bounded by $c$ if $s \geq c+1$, otherwise $l_1\leq s-1$. If, instead $s=1$ then $l_1=0$ and $l \in \{0, \ldots, c \}$.
\begin{widetext}
These probability can be explicitly constructed by iterating over $s$, starting from $s=1$ and up to a chosen maximum $s_{\rm max}$
\begin{equation} \label{eq:pils}
\begin{aligned}
\pi_{l,l_1} (s) &= Z_{\rm iter}^{-1}
 e^{\beta \left[ \mu + l \left ( \varepsilon - \frac{l+1}{2} \kappa_1 \right) \right] }
\frac{c!}{(c-l)!(l-l_1)!l_1!}
 \hat{F}_{l+1}^{c-l} \left[ (1 - p_B) \hat{R}_l \right]^{l-l_1} \\
 & \qquad \qquad \prod_{\alpha=1}^{l_1} \left[  
\sum_{s_\alpha} \left (  \sum_{m_\alpha = \theta(s_\alpha)}^{\min \{c, s_\alpha - 1 \}} \sum_{u_\alpha=m_\alpha}^c p_B \pi_{u_\alpha,m_\alpha} (s_\alpha) \, e^{-m_\alpha l \beta \kappa_2} \right) \right] \delta \left( 1+ \sum_{\alpha=1}^{l_1} s_\alpha = s \right) \, ,
\end{aligned}
\end{equation}
where $\theta(s_\alpha) = 0$ if $s_\alpha = 1$ and $\theta(s_\alpha) = 1$ if $s_\alpha>1$. One can  show that $\sum_{s=1}^{+ \infty} \sum_{l_1 = 0}^l \pi_{l,l_1} (s) = R_l$, provided that the normalization factor $Z_{\rm iter}$ is taken as in Eq.~\eqref{eq:cavity}.
The auxiliary function $\hat{F}_q,\hat{R}_q$ are defined in Eq.~\eqref{eq:aux}.
As explained above, setting $p_B=1$ describes geometrical clusters (in that case only the terms with $l=l_1$ are different from zero), and setting $p_B=1 - e^{-\beta \varepsilon/2}$ describes physical clusters.
These probabilities can be computed sequentially, because $\pi_{l,l_1}(s)$ are only functions of the cavity fields and of $\pi_{l^\prime,l_1^\prime}(s^\prime)$ for $s^\prime < s$. The computational cost, however, increases exponentially with $s_{\rm max}$.

Finally, the probability that a particle on a given site of the original lattice (with $c+1$ neighbors) belongs to a cluster of size $1<s<s_{\rm max} + 1$ reads:
\begin{equation} \label{eq:Pis}
\begin{aligned}
\Pi (s) =& Z_{\rm site}^{-1} \sum_{l_1=1}^{\min \{c+1, s - 1 \}} \sum_{l=l_1}^c 
e^{\beta \left[ \mu + l \left ( \varepsilon - \frac{l-1}{2} \kappa_1 \right) \right] }
\frac{(c+1)!}{(c+1-l)!(l-l_1)!l_1!}
\hat{F}_l^{c+1-l} \left[ (1 - p_B) \hat{R}_{l-1}\right]^{l-l_1}
\\
& \qquad  \qquad \prod_{\alpha=1}^{l_1} \left[ \sum_{s_\alpha}  \left ( \sum_{m_\alpha = \theta(s_\alpha)}^{\min \{c, s_\alpha - 1 \}} \sum_{u_\alpha=m_\alpha}^c p_B \pi_{u_\alpha,m_\alpha} (s_\alpha) \, e^{-m_\alpha (l-1) \beta \kappa_2} \right) \right] \delta \left( 1+ \sum_{\alpha=1}^{l_1} s_\alpha = s \right)
 \, .
\end{aligned}
\end{equation}
Again, it is possible to show that $\sum_{s=1}^{+ \infty} \Pi (s) = {\cal O} = \rho$, provided that $Z_{\rm site}$ is chosen as in Eq.~\eqref{eq:final}.
Using the fixed point from Eqs.~\eqref{eq:cavity}, Eqs.~\eqref{eq:pils} and~\eqref{eq:Pis} can be solved numerically up to $s_{\rm max}$. Note, however, that upon approaching the percolation transition (from low densities or high temperatures) $\Pi(s)$ develops power-law tails at large $s$,\cite{SA94} $\Pi(s) \sim s^{-3/2}$, so a diverging $s_{\rm max}$ must then be used to properly capture the distribution.
\end{widetext}

\subsection{Correlation function and linear stability} \label{sec:stability}

The density-density correlation function between two points at distance $r$ on the lattice can be  computed as the response to perturbing the chemical potential.\cite{RBMM04,MM09} Taking a given node $0$ as the origin, and numbering nodes on the tree at (Hamming) distance $r$ from it gives $\Omega(r) = (c+1) c^{r-1}$ neighbors. Applying an infinitesimal variation of the chemical potential on one of these nodes, exploiting the homogeneity of the high-temperature phase, and applying the fluctuation-dissipation theorem, provides the connected correlation function
\begin{equation} \label{eq:n0nr}
    \langle n_0 n_r \rangle_c = \frac{\partial \langle n_0 \rangle}{\partial (\beta \mu_r)} \, ,
\end{equation}
from which one immediately obtains the radial distribution function
\begin{equation}
    g(r) \equiv \frac{\langle n_0 n_r \rangle}{\rho^2} =  1 + \frac{1}{\rho^2 
    }\frac{\partial \langle n_0 \rangle}{\partial (\beta \mu_r)} \, .
\end{equation}
The correlation function is intimately related to the linear susceptibility of the homogeneous solution (which is the response to a global perturbation of $\mu$):
\begin{equation} \label{eq:chi}
    \chi \equiv \frac{\partial \rho}{\partial (\beta \mu)} = \rho(1-\rho) +  \sum_{r=1}^\infty \Omega(r) \langle n_0 n_r \rangle_c \, .
\end{equation}
Note that because the number of neighbors $\Omega(r)$ grows exponentially with distance, the series converges only if the density-density correlator decays exponentially over a correlation length $\xi$ such that $\xi < 1/\ln(c)$. If instead $\xi \ge 1/\ln(c)$ the susceptibility diverges and the homogeneous solution is unstable.
Because $\langle n_0 \rangle = {\cal O}_0 /({\cal E}_0+{\cal O}_0)$ and ${\cal E}_0$ and ${\cal O}_0$ are only functions of the cavity fields on the neighbors of $0$ (at distance $r=1$), one can use the chain rule to express the derivative of Eq.~\eqref{eq:n0nr} as:
\begin{equation} \label{eq:chain}
   \langle n_0 n_r \rangle_c =  \frac{\partial \langle n_0 \rangle}{\partial \vec{\varphi}_1}
   \frac{\partial \vec{\varphi}_1}{\partial \vec{\varphi}_2}  \ldots  \frac{\partial \vec{\varphi}_{r-1}}{\partial \vec{\varphi}_r} \frac{\partial \vec{\varphi}_{r}}{\partial (\beta \mu_r)} \, ,
\end{equation}
where $\vec{\varphi}_n$ denotes the set of all $4(c+1)$ cavity fields on the $n$-th generation of the tree on the branch $(0 \leftrightarrow 1 \leftrightarrow 2 \leftrightarrow \ldots \leftrightarrow r-1 \leftrightarrow r)$ that connects site $0$ with site $r$. Because in the high-temperature phase all these fields are equal and independent of $r$, we introduce the $4(c+1) \times 4(c+1)$ Jacobian matrix
\begin{equation} \label{eq:J}
    {\cal J} = 
    \left . \frac{\partial \varphi_{m}^{(i \to j_0)}}{\partial \psi_{m^\prime}^{(j_1 \to i)}} \right \vert_{\rm h.s.} \, ,
\end{equation}
where $(\varphi_m,\psi_m) = \{ E_m,F_m,O_m,R_m \}$. As detailed in Appendix~\ref{app:jacobian}, the matrix elements of this Jacobian are computed by taking the derivatives of Eqs.~\eqref{eq:cavity_inhomogeneous} with respect to the cavity fields on one of the neighbors (say $j_1$) and evaluating the resulting expressions when all the cavity fields take the values corresponding to the homogeneous solution of Eqs.~\eqref{eq:cavity} (see Eq.~\eqref{eq:Japp}).
Denoting $\lambda_{\rm max}$ the eigenvalue of largest modulus of ${\cal J}$, from~\eqref{eq:chain} one immediately obtains that at large distance $\langle n_0 n_r \rangle_c \sim \lambda_{\rm max}^r$ and $\xi^{-1} = -\ln( \vert \lambda_{\rm max} \vert)$. Because $\Omega(r) \approx c^r$, from Eq.~\eqref{eq:chi} the stability criterion simply reads $c \vert \lambda_{\rm max} \vert \le 1$. When $c \vert \lambda_{\rm max} \vert > 1$, the paramagnetic solution is unstable to perturbations that are homogeneous within a generation. This instability is towards either gas-liquid coexistence (ferromagnetic ordering) or periodic microphases, in which successive layers carry different fields. Such analysis can therefore be used to determine the spinodal lines and the critical point of the gas-liquid coexistence (Figs.~\ref{fig:PD}a and~\ref{fig:PD}b) as well as the order-disorder transition (ODT) of periodic microphases. (At small connectivity (Fig.~\ref{fig:PD}c), such as $c=2$ and $c=3$, the transition is concomitant with the loss of local stability of the high-temperature phase.)

Within the same framework, one can also consider an instability associated with the divergence of the non-linear (spin-glass) susceptibility
\begin{equation} \label{eq:chiSG}
    \chi_{\rm SG} = \rho^2(1-\rho)^2 + \sum_{r=1}^\infty \Omega(r) \langle n_0 n_r \rangle_c^2 \, .
\end{equation}
When $\chi_{\rm SG}$ diverges a replica-symmetry breaking\cite{MP87} instability occurs, which appears as a widening of the variance of the cavity fields under the recursion relations in Eqs.~\eqref{eq:cavity_inhomogeneous}.\cite{MM09,BM02,CTCC03b,RBMM04,KTZ08,CFT20} Such spin-glass instability thus takes place when $c  \lambda_{\rm max}^2 \ge 1$.
The eigenvalue $\lambda_{\rm max}$ is the same as above because the transfer matrix is simply the square of the
Jacobian defined in Eq.~\eqref{eq:J}. Although this condition is always weaker than for the modulation instability, $c \vert \lambda_{\rm max} \vert > 1$, it is putatively relevant for RRGs, as discussed in Sec.~\ref{sec:inhomogeneous}. The lines of the $\chi_\mathrm{SG}$ divergence are thus included in the phase diagrams of Figs.~\ref{fig:TcTl} and~\ref{fig:PD}c. Note, however, that the actual glass transition may well take place at higher temperatures.\cite{MP87}

\section{Inhomogeneous solutions} \label{sec:inhomogeneous}
As anticipated above, at low temperatures and for strong enough repulsion the homogeneous solution becomes linearly unstable towards microphase ordering in which different generations of the recursive equations carry different cavity fields. 
In order to study this inhomogeneous regime, in which the equilibrium phase is no longer translationally invariant, one seeks solutions of the cavity equations for which the cavity fields can vary periodically from one lattice site to another.

Such solutions have already been studied in the context of SALR models on loop-less Cayley trees in Refs.~[\onlinecite{Va81,IT83,YOS85,MCA85,SC86,RAU14}], and a rich variety of low-temperature modulated phases have been identified for different model parameters. While on a Cayley tree these phases are all characterized by radial density profiles that vary periodically on subsequent concentric ring generations, such periodic solutions are incompatible with the RRGs used to describe the homogeneous phases because loops frustrate periodic ordering. An alternative route is to design suitable tree-like (boundary-less) random graphs that are compatible with the specific ordering of interest. For succinctness, we here only consider the lamellar phase that establishes for $\rho\sim 1/2$, but the same procedure could be adapted and extended to other low-temperature modulated structures, such as the columnar and crystal-cluster phases.

Consider $L$ independent realizations of RRGs of connectivity $c-1$ and place them at consecutive positions along the $x$ axis. Each node of the RRG at position $x$ is then connected to two randomly chosen nodes of the RRGs at coordinates $x-1$ and  $x+1$, in such a way that the total connectivity of the graph remains $c+1$. 
This construction results in a layered version of the RRG that is anisotropic in the $x$ direction yet remains locally tree-like, because two sites connected by an edge within a given layer at coordinate $x$ are not connected (with high probability in the thermodynamic limit) on the other layers and short loops are rare.

Imposing that in the lamellar phase the system is translationally invariant along the transverse direction simplifies Eqs.~\eqref{eq:cavity_inhomogeneous} to a system of $12 L (c+1)$ coupled non-linear algebraic equations (which we do not write down explicitly for the sake of concision) for the set of cavity fields that depend only on the $x$-coordinate. The fixed point of these equations, which can be found numerically by iteration, gives access to all the thermodynamic observables of the striped phase (such as the density profile, the average energy and entropy, the free energy, the correlation function, \emph{etc}.) by adapting Eqs.~\eqref{eq:final_inhomogenous} and~\eqref{eq:freegraph} to the particular graph geometry.

Two key technical difficulties, however, then arise. The first is associated with the infinite number of possible lamellar solutions, related by global translations of the density profile. In order to pin one of those solutions and to converge numerically to a single fixed point, two fully occupied layers are placed at coordinates $x=0$ and $x=L+1$, and the recursion relations only solved for $1 \le x \le L$.
The second is related to the commensurability constraint for the number of layers with equilibrium periodicity $\lambda_\ell$, which is not known a priori and depends on model parameters and thermodynamic conditions. The recursion equations must therefore be solved for several values of $L$ in order to identify the striped solution with the lowest free energy. (The same problem is observed, for instance, in numerical simulations and is similarly solved.\cite{ZC10,ZC11,ZZC16})

It is interesting to note that in the $Z_2$ symmetric case ($\rho=1/2$) and close to the critical point $T \lesssim T_c(\kappa)$, the equilibrium properties of the lamellar phase found on the anisotropic layered RRG above are the same as for the periodic antiphase found on concentric shells of the Cayley tree in Ref.~[\onlinecite{SC86}]. Neither the value of $T_c(\kappa)$ nor the equilibrium periodicity $\lambda_\ell$ are affected by the specific geometry of the underlying lattice.

\section{Discussion}
\label{sec:discussion}
\begin{figure*}
	\includegraphics[width=0.40\textwidth]{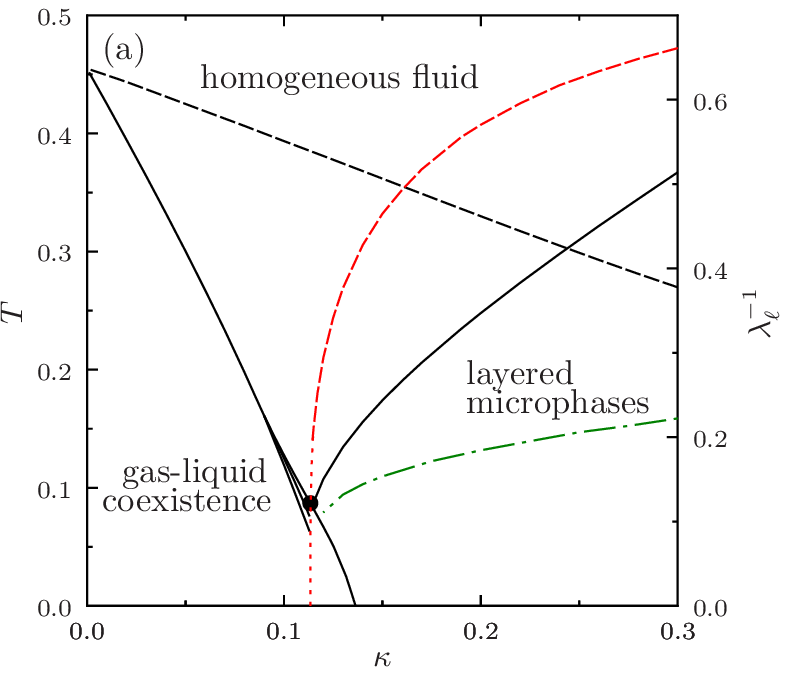} \hspace{1cm}
	\includegraphics[width=0.40\textwidth]{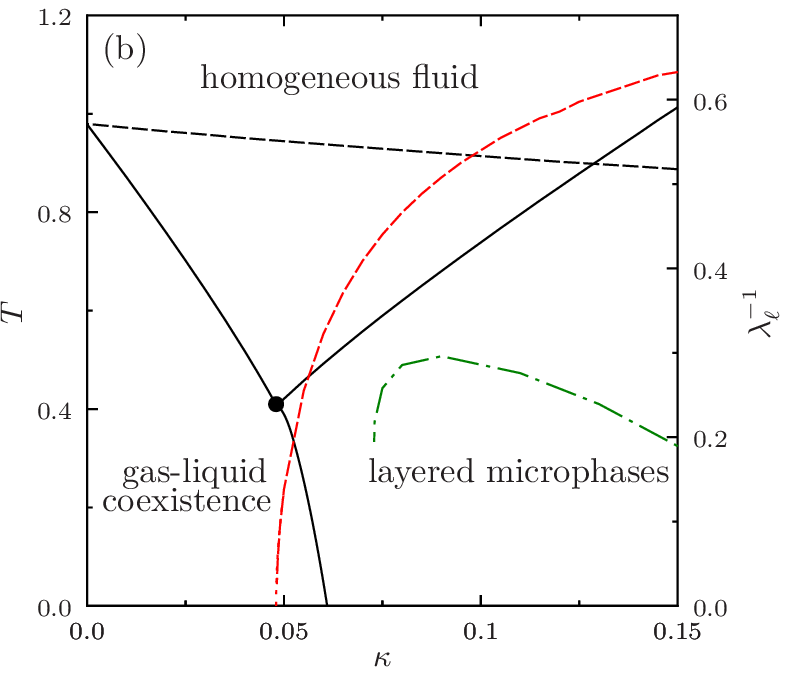}
	\caption{Phase diagrams at $\rho=1/2$ (with $Z_2$ symmetry for $\mu$ given by Eq.~\eqref{eq:mu05}) for (a) $c+1=3$ and (b) $c+1=5$. At small $\kappa$, the critical line separates the high-temperature fluid (paramagnetic) phase from low-temperature gas-liquid coexistence (ferromagnetic phase), and is initially in the Ising universality class. In (a), however, a line of first-order transitions emerges before reaching the Lifshitz point at $\kappa_L$ (black dot). For $\kappa>\kappa_L$, the transition to the ordered lamellar phase is in the XY universality class. The multicritical Lifshitz point at (a) ($\kappa_L=0.1134$, $T_c(\kappa_L)=0.0881$) and (b) ($\kappa_L=0.0481$, $T_c(\kappa_L)=0.407$) separates one critical regime from the other. Its extension to $T<T_c(\kappa_L)$ smoothly connects to the energetic onset of modulation at $T=0$ given by Eq.~\eqref{eq:T0mod}. The layer thickness $\lambda_\ell$ along the critical line (dashed line, right axis) diverges at the Lifshitz point. Because the nature of transitions meeting at $\kappa_L$ differs, the critical scaling of $\lambda_\ell$ (dotted red lines) also differs (see Eq.~\eqref{eq:lambdaL}). The physical percolation threshold (dashed lines) coincides with the critical point only for the standard lattice-gas ($\kappa=0$) model, lying first above and then below $T_c$ as $\kappa$ increases. The lines of $\chi_\mathrm{SG}$ divergence (dashed-dotted green lines) offer a lower bound on the point at which the homogeneous solution---if somehow continued in through the layered microphases---eventually becomes unstable towards glass formation.}
    \label{fig:TcTl}
\end{figure*}

\begin{figure*}
	 \hspace{-0.4cm}\includegraphics[width=0.33\textwidth]{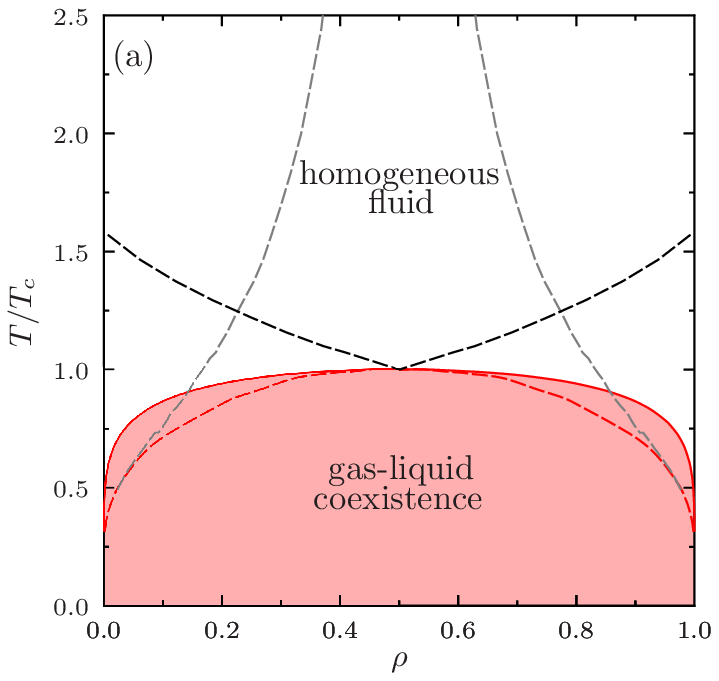}	 \hspace{-0.4cm}
	\includegraphics[width=0.33\textwidth]{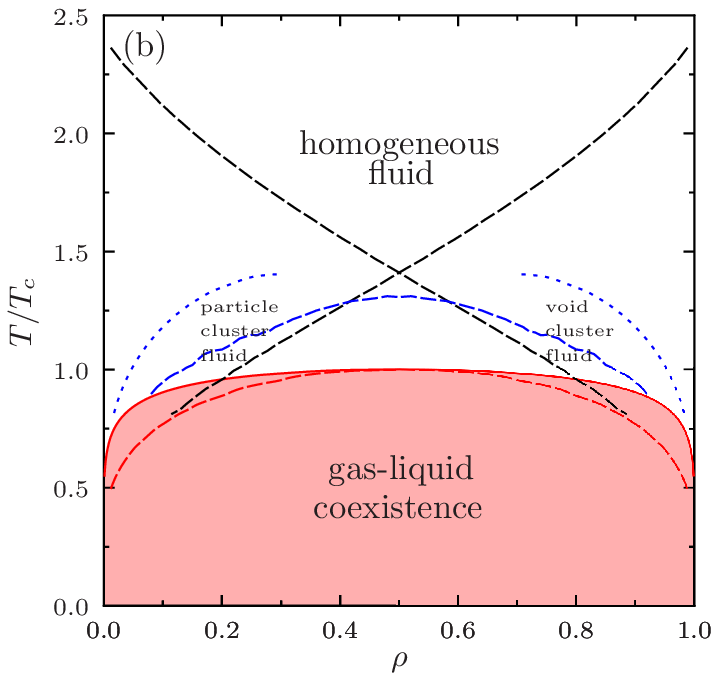} \hspace{-0.4cm}
	\includegraphics[width=0.33\textwidth]{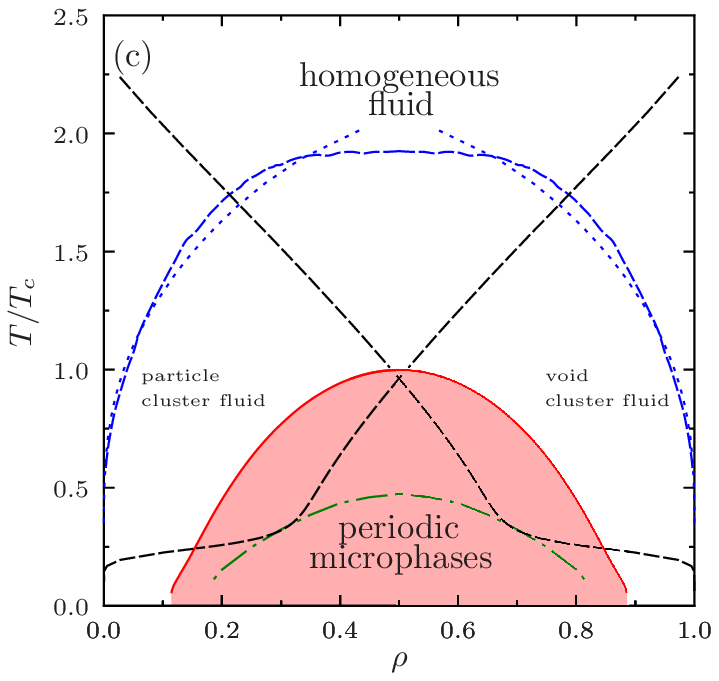}
	\caption{Temperature-density phase diagram for $c+1=3$ with (a) $\kappa=0$ ($T_c(0) = 0.455$), (b) $\kappa=0.05$ ($T_c(0.05) = 0.301$) and (c) $\kappa=0.25$ ($T_c(0.25)= 0.311$). Panels (a) and (b) have $\kappa<\kappa_L=0.1134$, while (c) has $\kappa>\kappa_L$. The ordered phase (red region) thus corresponds in (a) and (b) to gas-liquid coexistence---with a spinodal instability (red dashed line)---and in (c) to periodic microphases. The geometrical (dashed gray lines in (a)) and physical (dashed black lines, CK for $p_B = 1 - e^{-\beta \varepsilon/2}$) percolation lines of clusters and voids coincide at low $T$, as $p_B\to1$, but separate as $T$ increases. The geometrical percolation line then tends to the random percolation threshold $\rho=1/c=1/2$ in the $T \to \infty$ limit, while its physical counterpart terminates at a finite $T$. The blue dashed lines in (b) and (c) identify the locus of the heat capacity maximum, as in Fig.~\ref{fig:C}. In panel (c) it coincides with the the physical clustering crossover (defined as the density at which $\Pi(s=1) = 4 \Pi(s=2)/3$, dotted blue lines). The $\chi_\mathrm{SG}$ divergence (green dashed-dotted line) in (c) indicates where the homogeneous solution becomes unstable towards replica-symmetry breaking, thus providing a lower bound for the fluid phase to undergo a glass transition.}
    \label{fig:PD}
\end{figure*}

\begin{figure}
	\hspace{-0.2cm}\includegraphics[height=0.495\columnwidth]{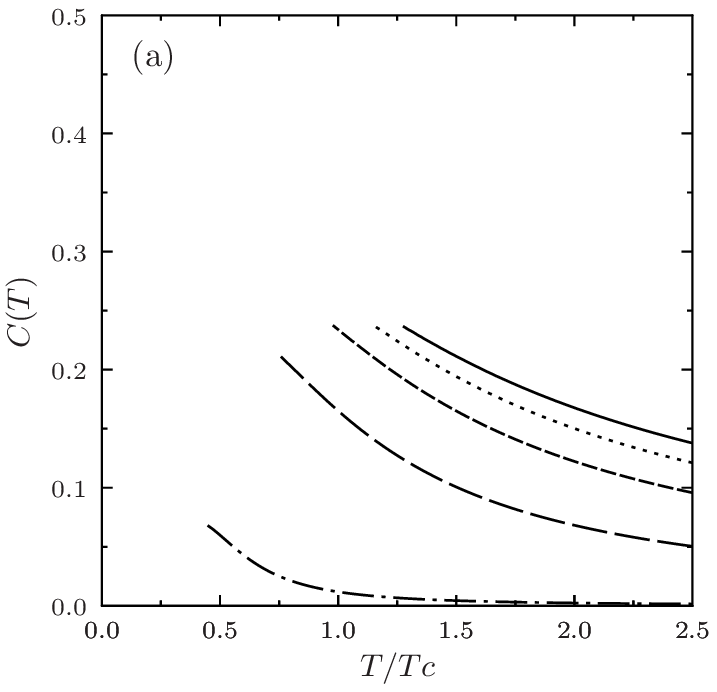}\hspace{-0.2cm}
	\includegraphics[height=0.495\columnwidth]{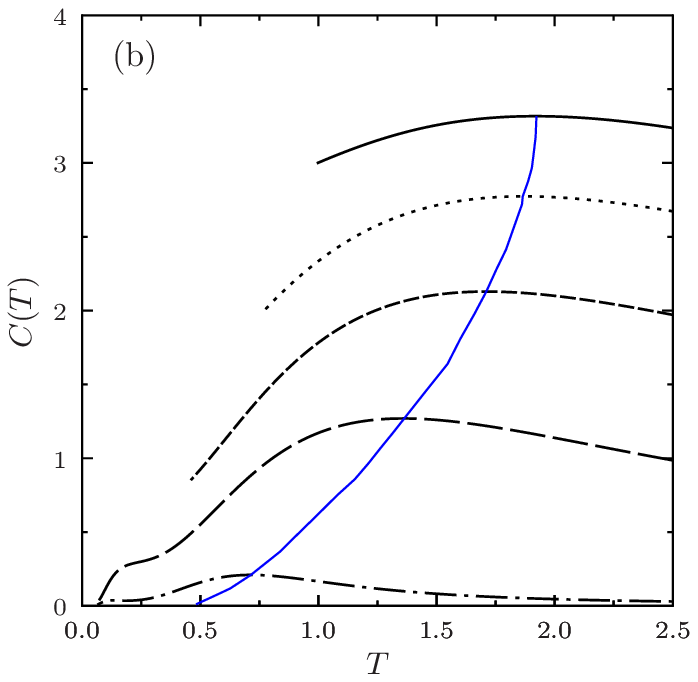}
	\caption{Evolution of the specific heat $C(T)=\partial e / \partial T$ in the high-temperature phase for (a) $\kappa=0$ and (b) $\kappa=0.25$ at $\rho=0.01$, 0.1, 0.2, 0.3, 0.5, from bottom to top. For the standard lattice-gas case in (a), $C$ increases monotonically as $T$ decreases, up to the gas-liquid coexistence line. By contrast, in the clustering case (b) $C$ is maximal (red curve) well above the microphase ordering temperature (see Fig.~\ref{fig:PD}). Note the difference in scale between the two panels and that $C$ here does not diverge at $T_c$, because the associated critical exponent $\alpha$ generically vanishes in mean-field systems.}
    \label{fig:C}
\end{figure}

\begin{figure}
	\includegraphics[width=0.23\textwidth]{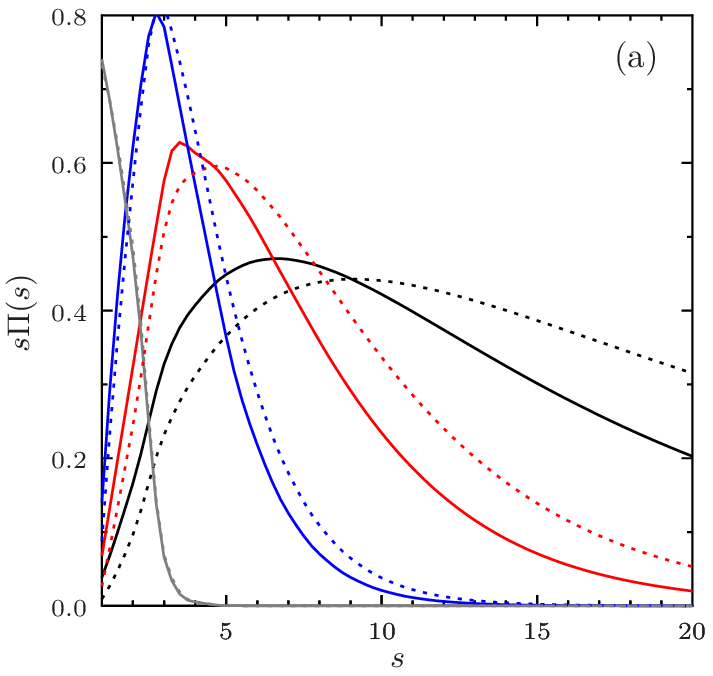}
	\includegraphics[width=0.23\textwidth]{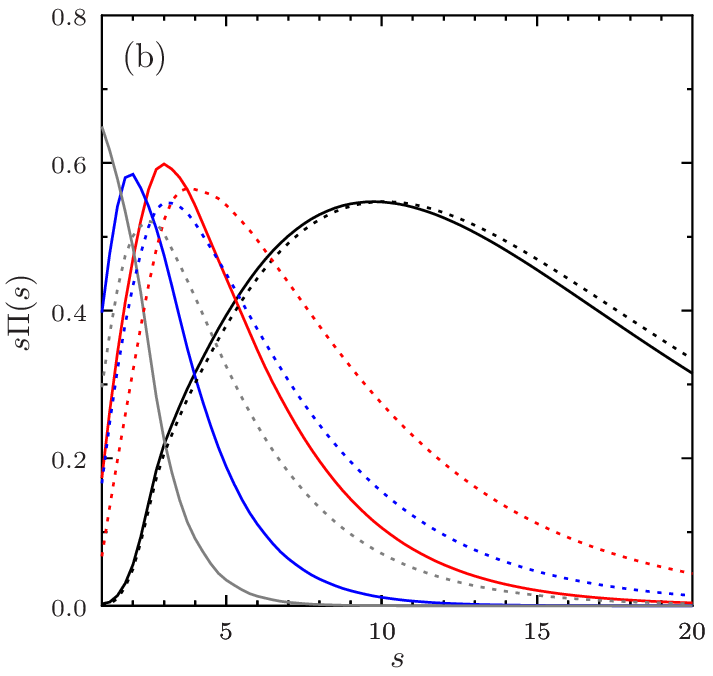}\\
	\includegraphics[width=0.23\textwidth]{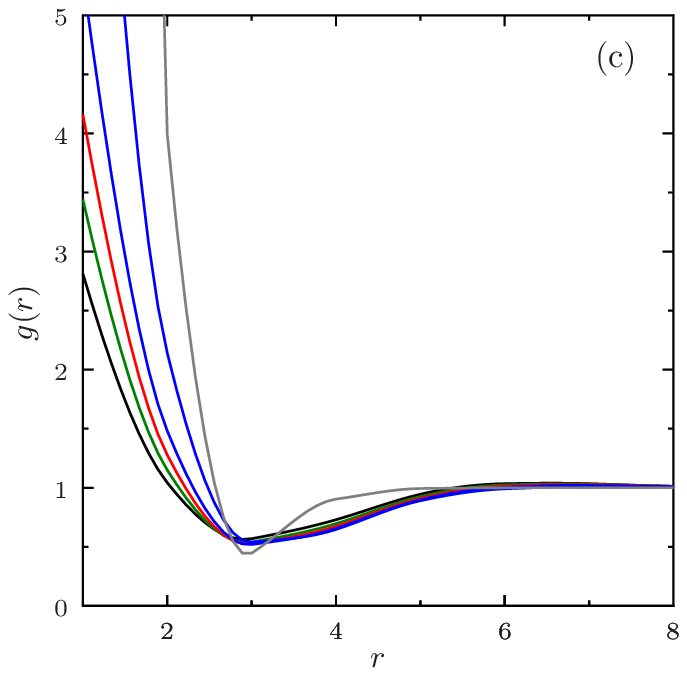}
	\includegraphics[width=0.23\textwidth]{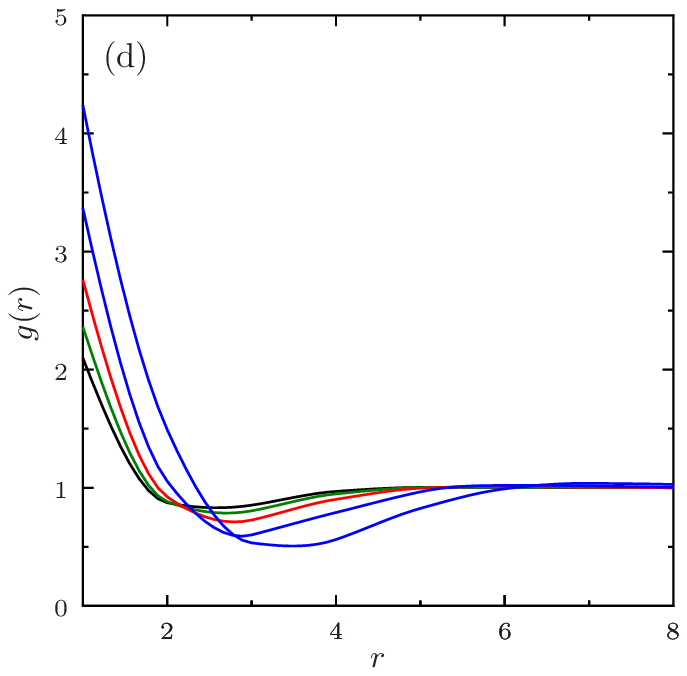}
	\caption{Cluster size distribution $s \Pi(s)$ in the high-temperature phase for $c+1=3$, $\kappa=0.25$, and (a) at $T = 0.2 \approx 2 T_L/3$ for $\rho =0.001$, $0.06$, $0.12$, $0.16$, $0.254$, and (b) at $\rho=0.18$ for $T=0.9$, $0.7$, $0.5$, $0.3$, and $0.1125$,  upon approaching the instability toward microphase ordering. Continuous lines correspond to physical ($p_B=1-e^{-\beta \varepsilon/2}$) clusters, while dashed lines to geometrical clusters ($p_B=1$). (c) and (d) show the essentially featureless pair correlation function $g(r)$ for the same temperatures and densities as for the top panels.}
    \label{fig:clusters}
\end{figure}
Using the methods described in the previous sections, it is possible to analyze the phase diagram and structural feature of the SALR model defined by Eq.~\eqref{eq:H}. We first determine the onset of periodic microphases, in order to discern that regime from standard liquid-gas coexistence.  For simplicity we specifically consider the $Z_2$ symmetric case with $\rho=1/2$ (see Fig.~\ref{fig:TcTl}) by setting the chemical potential as in Eq.~\eqref{eq:mu05}. In this case, the ODT temperature is maximal for a given $\kappa$ and, by symmetry, the ordered phase is lamellar. As expected, at weak $\kappa$ there is no modulated solution of the recursion relations, and along the critical line, $T_c(\kappa)$, the order-disorder transition remains in the Ising universality class with $T_c$ simply depressed.
At large $\kappa$, however, a lamellar solution with a periodic density profile, $\rho(x) \approx \rho_\ell \sin( 2 \pi x / \lambda_\ell + \phi)$, develops below a critical temperature $T_c(\kappa)$ (see Sec.~\ref{sec:inhomogeneous}). For the (mean-field) model considered here, the ODT to this lamellar phase is continuous, with $\rho_\ell$ vanishing at $T_c(\kappa)$ upon increasing the temperature as $\rho_\ell \sim (T_c(\kappa)-T)^{1/2}$ and with $\lambda_\ell$ finite, except at the tricritical point, as discussed below. (In three dimensions, the continuous ODT is replaced by a Brazovskii's fluctuation-induced weakly first-order transition \cite{B75,FH87,GKTV02,TC06,TC07,matsen1996unifying,ZC17}). By symmetry, one expects the transition to be in the XY universality class, which we confirm by noting that the linear susceptibility obtained from Eq.~\eqref{eq:chi} diverges for $T<T_c(\kappa)$ in the whole lamellar phase. In other words, the pair correlation $\Omega(r) \langle n_0 n_r \rangle_c$ oscillates with a constant amplitude that does not decay with $r$, due to a massless Goldstone mode with long-range correlation associated with global translations of the density profile along $x$.

Interestingly, the nature of the regime intermediate between small and large $\kappa$ depends on lattice connectivity. For $c\geq 4$, a standard Lifshitz tricritical point at ($\kappa_L$, $T_c(\kappa_L)$) separates the two regimes. For $2 \le c \le 3$, however, the Ising-like critical point first transforms into a first-order transition before the multicritical point is reached. The onset of the gas-liquid transition then proceeds discontinuously. In this case the Lifshitz point appears at the termination of the spinodal line of the paramagnetic phase (Fig.~\ref{fig:TcTl}). 
In addition, while for $c \ge 4$ the homogeneous solution remains locally (meta)stable below $T_c$,  for $2 \le c \le 3$ the transition to the layered microphase coincides with its loss of linear stability. To the best of our knowledge, these peculiarities have not been noted before. Because they have no notable impact on the microphase-forming regime beyond $\kappa_L$, however, we do not here explore the matter further.

At $T<T_c(\kappa_L)$, a similar distinction can be drawn between gas-liquid coexistence and layered microphases. At $T=0$ the transition is straightforwardly given by Eq.~\eqref{eq:T0mod}, while at finite $T$, the onset of modulation can be determined from the inhomogeneous solutions described in Sec.~\ref{sec:inhomogeneous}. Similarly, this treatment can be used to determine the equilibrium periodicity $\lambda_\ell$ along the $T_c(\kappa>\kappa_L)$ line, which diverges at $\kappa_L$ as
\begin{equation} \label{eq:lambdaL}
\begin{aligned}
\lambda_\ell &
\simeq \sqrt{2} \pi^{\frac{3}{4}} (\sqrt{\kappa} - \sqrt{\kappa_L})^{-\frac{1}{2}} \propto (\kappa-\kappa_L)^{-\frac{1}{2}} & c \ge 4 \, ,\\
\lambda_\ell & \propto (\kappa-\kappa_L)^{-\frac{1}{4}} & 2 \le c \le 3 \, ,
\end{aligned}
\end{equation}
depending on the nature of the multicritical point.\cite{TC07}

Figure~\ref{fig:TcTl} also shows how the physical percolation threshold, $T_{\rm perc}(\kappa)$  for $\rho=1/2$ evolves. (For $\rho=1/2$, clusters and voids geometrically percolate for all $T$ and $c$, because random percolation threshold, \emph{i.e.}, for $T \to \infty$, on tree-like graphs is $\rho_{\rm perc} = 1/c \le 1/2$.)  In the standard lattice gas ($\kappa=0$) model, physical percolation (with $p_B=1-e^{-\varepsilon/2}$) coincides with the critical point of gas-liquid coexistence.\cite{CK80} This feature underlies the efficient sampling of configurations through Fortuin-Kasteleyn-cluster--based moves (which coincide with CK physical percolation at the critical point), such as in the Swendsen-Wang algorithm.\cite{SW87} Increasing $\kappa$, however, differently affects the percolation threshold and the critical temperature, and hence $T_{\rm perc} (\kappa)$ is generically different from $T_c (\kappa)$ for $\kappa>0$. In particular, at small $\kappa$, Fortuin-Kasteley clusters percolate well above $T_c$. This dissociation opens up a gel-like region, in which both particles and voids percolate, and within which the system exhibits a finite macroscopic elastic response.\cite{ZZBRDX,XMG04,DdC02}
This dissociation further suggests that Fortuin-Kasteley-cluster--based sampling algorithms then cannot be used for meaningful structural relaxation, as has indeed been reported for various frustrated models.\cite{LH91,CFNSC96,KHFP08,JSGS13,RD17,FHY20} At large $\kappa$, by contrast, $T_{\rm perc} < T_c$ and therefore no equilibrium bicontinuous fluid structure forms. A potentially long-lived metastable gel-like  structure might then only be achievable if periodic microphase ordering is dynamically avoided upon supercooling below $T_c$ (such as by considering lattice geometries that inhibit the development of long-range periodic order). Note that FK cluster-based algorithms are then nevertheless inefficient because they do not relax the relevant diverging spatial correlations in the vicinity of $T_c$.

Further supercooling the fluid phase below $T_c$ (while avoiding periodic microphase ordering), eventually makes it unstable towards replica symmetry breaking, as indicated by the line of diverging $\chi_\mathrm{SG}$ from Eq.~\eqref{eq:chiSG} (see Fig.~\ref{fig:TcTl}).\cite{MP87} 
This finding implies the existence of a glassy phase at low enough temperature and strong enough repulsion, similarly to what was originally suggested in the context of Ginzburg-Landau $\phi^4$ models with Coulomb repulsion.\cite{SW00,HSW01,GTV02,GTV02b} (The chaotic phases associated with strange attractors found in the family of SALR models described by Eq.~\eqref{eq:H} on the Cayley tree in Ref.~[\onlinecite{YOS85}] might also be a manifestation of this glass phase.) A more careful study of the glassy regime and of the glass transition is, however, beyond the reach of the simple replica symmetric treatment considered here. Upon lowering the temperature, one could indeed either have a continuous spin-glass transition right at the instability point,\cite{CFT20} or (more likely) a random first-order glass transition beforehand, as reported for certain continuous space SALR models.\cite{SW00,HSW01,GTV02,GTV02b,TC06,TC07,campbell2005dynamical,toledano2009colloidal} In order to distinguish between these two scenarios and to determine the onset of glassiness one would need to implement a (much more involved) 1RSB solution of the model.\cite{MP01,BM02,CTCC03,CTCC03b,T07,RBMM04,KTZ08,CFT20} The question of whether or not a stable glassy phase can exist above $T_c$ in a SALR mean-field model is thus left for future investigations. 

Having clarified the role of connectivity and of $\kappa$ on the phase behavior of the $Z_2$ symmetric case, we next consider the overall phase structure in the $(\rho-T)$ plane for different $\kappa$, above and below $\kappa_L$. For the rest of this study, we set the total connectivity $c+1=3$, without loss of generality as long as we stay either well above or well below $\kappa_L$. This particular connectivity is chosen largely for numerical convenience, but also because it gives rise to slightly larger clusters than larger $c$.

As reference, we first consider the case $\kappa=0$, for which Eq.~\eqref{eq:H} reduces to the standard lattice-gas Hamiltonian.  As expected---and as can be seen in Fig.~\ref{fig:PD}(a)---the homogeneous (on average) fluid phase then gives way to gas-liquid coexistence for $T<T_c$. The two percolation lines---geometrical and physical---for both particle and voids are naturally equivalent at low temperatures (because $p_B \to 1$ for $T \to 0$), but separate as $T$ increases. As expected, the geometrical percolation line approaches $1/c=1/2$ as $T\to\infty$, and does not capture much of the relevant physics of the fluid phase. (In subsequent cases, the geometrical percolation line is omitted for clarity.) The physical (cluster and void) percolation lines, by contrast, closely follow the spinodal instability line,  cross at the critical point, and terminate at a finite $T$.\cite{CK80} The supercritical fluid regime otherwise exhibits no obvious inhomogeneity nor thermal anomaly.
For $\kappa>0$ yet still well below the Lifshitz point, interesting structural features emerge in the fluid phase, as for $\kappa=0.05$ in Fig.~\ref{fig:PD}(b). In particular, even though the physical percolation lines still coincide with the spinodal at low $T$, they eventually detach and cross well above the critical point (as shown in Fig.~\ref{fig:TcTl}). A region in which both particles and voids percolate thus open up,  resulting in a bicontinuous fluid structure morphologically resembling a gel. 

For frustration well beyond the Lifshitz point, such as $\kappa=0.25$ in Fig.~\ref{fig:PD}(c), the model becomes structurally even richer. Most obviously, gas-liquid coexistence transforms into periodic microphases. By contrast to the envelope of the former, that of the latter---identified from the linear instability of the homogeneous solution as in Sec.~\ref{sec:stability}---does not extend to a vanishing density at low temperatures. It instead terminates at a finite $\rho \sim 0.1$ (or $\rho \sim 0.9$ at high density). As a result, the fluid phase survives over a broad density range down to $T=0$. Because this regime lies below the physical percolation line, the low-density fluid is then a percolated, mechanically rigid network supported by long-lived bonds between nearest-neighbor particles.\cite{TPCSW01,de2006columnar,sciortino2005one} The reduction of the density spread of periodic microphases can be made even more pronounced as $\kappa\to\kappa_L$. Hence SALR interactions lead to a complementary---and possibly more experimentally accessible---route for producing a stable low-density gel-like structure to that proposed in Refs.~[\onlinecite{DK07,DK07b,BLTZS06,ZSBMTS06,SLN06}] through decreasing the average particle coordination number.
Because ordered microphases are suppressed on RRGs, the physical percolation lines can also be followed within the linear instability regime of the fluid phase. For the particular $\kappa$ considered, we find that these lines cross below $T_c$ (Fig.~\ref{fig:TcTl}). Hence for this (and larger) $\kappa$, no equilibrium bicontinuous fluid structure forms. A potentially long-lived metastable such structure might nevertheless be achievable if microphase ordering is dynamically avoided, and then gets frozen in once the system undergoes a glass transition (lower bounded by the line of diverging $\chi_\mathrm{SG}$).

For both $\kappa=0.05$ and $\kappa=0.25$, the heat capacity peaks within the fluid regime (see Fig.~\ref{fig:C}). (For $\kappa=0$, no such peak is found.) Although this peak is somewhat broad, it narrows steadily as temperature decreases. In certain models, this feature has been identified with the onset of clustering, whose trend it generally follows.\cite{IR06,SCK16,PBC19,frantz1995magic} In order to assess this relationship, we separately consider the clustering properties of Eq.~\eqref{eq:H} as detailed in Sec.~\ref{sec:clusters}. Note that for the sake of succinctness, we here specialize to the case $\kappa=0.25$, but similar observations could be made for other $\kappa>\kappa_L$. Figure~\ref{fig:clusters} shows the evolution of the cluster size distribution $s \Pi (s)$, which is proportional to the number of particles belonging to a cluster of size $s$, with density and temperature, with $\Pi(s)$ normalized to $1$. (In order for thermally relevant clusters alone to be considered, a CK-like probabilistic treatment is applied, weighting interparticle bonds by their Boltzmann weight, $p_B = 1 - e^{-\beta \varepsilon/2}$.) At high temperatures and low densities, the distribution is dominated by single particles, as can be seen by the peak at $s=1$ accompanied by a simple exponential tail for $n=2$ (dimers), $n=3$, \emph{etc}. This behavior is akin to that of simple fluids. Upon lowering $T$ or increasing $\rho$, however, the cluster size distribution qualitatively changes. A second peak appears and then dominate as isolated particles rarefy. This crossover is smooth yet crisp. Further evolving $T$ and $\rho$ then steadily increases the mode of that second peak as well as its width, until either the cluster fluid becomes unstable to ordering or physical percolation is reached. Clustering being a smooth crossover, different observables locate its onset at slightly different conditions. In order to compare its position with that of the heat capacity peak, we therefore have a certain freedom of choice. Here, we phenomenologically define the onset of clustering as the density (or temperature) at which $\Pi(s= 1) = 4 \Pi(s= 2) /3$. This condition is intermediate between requiring that $\Pi(s)$ develops a peak at a value of $s$ larger than $1$, \emph{i.e.}, $\Pi(s=1) = \Pi(s=2)$, and that $s \Pi(s)$ develops a peak for $s>1$, i.e. $\Pi(s=1) = 2 \Pi(s=2)$.
Although this choice is largely arbitrary, it remains nonetheless constrained by the relatively small cluster sizes observed on RRGs (5-10 particles), compared to those reported for comparable interactions in real space (20 particles and more).\cite{ZC21} The resulting curve closely follows the heat capacity peak, at least up to $\rho\approx 1/3$ (Fig.~\ref{fig:PD}). This behavior contrasts with that of the equation of state of off-lattice systems.\cite{ZZC16,ZC17,SPG17,HC18,ZC21} The (osmotic) pressure signature of clustering then only persists to a finite density, after peeling off of the heat capacity curve. Our lattice model, however, does provide further insight into the origin of this effect, because its equation of state exhibit no such feature and the microscopic reason for its absence remains unclear. Another limitation of our approach is that the cluster-shape instability---from spherical to elongated---recently reported in a continuum space SALR model\cite{ZC21,sciortino2005one,de2006columnar} cannot be directly assessed. A spatially inhomogeneous version of the cluster equations would then be needed, at a markedly larger analytic and computational cost. 

Interestingly, cluster aggregation leaves no obvious trace on the pair correlation function, $g(r)$, at least at short distances. Pair correlations evolve smoothly in a way that is essentially indistinguishable from what happens in a simple liquid. Although $g(r)$ exhibits oscillations precursor to periodic ordering on the characteristic $\lambda_\ell$ scale, as $g(r) \sim e^{-r / \xi} \sin (2 \pi r / \lambda_\ell)$,\cite{TC06,TC07,HSW01} cluster formation does not correlate with these oscillations, which persist even at high temperature and small density where $\Pi(s)$ has a maximum in $s=1$ followed by a simple exponential decay. This behavior contrasts with the local ordering signature carefully teased out for certain systems.\cite{bomont2020local} At the very least, this observation suggests that such a correlation is not essential to cluster formation.

\section{Conclusion}
\label{sec:conclusion}
In this work, we have solved a simple lattice SALR model on a RRG. Although in many ways different from a real-space model, let alone an experimental system, this exactly-solvable model recapitulates many of the key structural features observed in the high-temperature fluid phase of a variety of more elaborate and realistic models.\cite{ZC21} This correspondence suggests that the strong universality of periodic microphase formation extends to the disordered regime as well. 

This work also suggests that several directions remain to pursue to capture the key physics of even elementary SALR models. First, the emergence of a first-order transition ahead of the Lifshitz point at low connectivity appears to be relatively robust, yet its physical origin remains somewhat nebulous. Second, reconciling the differential evolution of the standard physical percolation threshold and the critical point could significantly improve the design of cluster-based Monte Carlo sampling. Third, resolving the interplay between structure and dynamics for disordered microphases could guide the experimental design of microphase self-assembly more generally. Finally, studying the glass transition $T_g$ of SALR models through a RSB treatment might reveal that $T_g > T_c$ for some $\kappa$, which could be of considerable theoretical and experimental interest for the study of amorphous solids.

\begin{acknowledgements}
We thank Yi Hu, Ye Liang and Mingyuan Zheng for stimulating discussions as well as for sharing results ahead of publication. PC was supported by a the National Science Foundation Grant No.~DMR-1749374. Data relevant to this work have been archived and can be accessed at the Duke Digital Repository.\cite{mfdata}
\end{acknowledgements}

\appendix

\section{Computation of the Jacobian} \label{app:jacobian}

In order to compute the Jacobian associated with the recursion relations, we introduce eight $(c+1)\times(c+1)$ square matrices whose elements are the derivatives of the numerators of the right-hand side of the recursion relations in Eq.~\eqref{eq:cavity_inhomogeneous} with respect to the cavity fields of neighbor $j_1$ divided by the normalization $Z_{\rm iter}$. At the end of the calculation, all the cavity fields are evaluated in the homogeneous solution (h.s.) of Eq.~\eqref{eq:cavity}, thus giving
\begin{equation}
\begin{aligned}
    {\cal J}^{(EE)}_{l,l^\prime} & =\frac{(c-l) e^{-\beta l l^\prime \kappa_2} E_l}{c \hat{E}_l} \, , \\
    {\cal J}^{(EO)}_{l,l^\prime} & =\frac{l e^{-\beta (l-1) l^\prime \kappa_2} E_l}{c \hat{O}_{l-1}} \, , \\
    {\cal J}^{(FE)}_{l,l^\prime} & =\frac{(c-l) e^{-\beta (l+1) l^\prime \kappa_2} F_l}{c \hat{E}_{l+1}} \, , \\
    {\cal J}^{(FO)}_{l,l^\prime} & =\frac{l e^{\beta l l^\prime \kappa_2} F_l}{c \hat{O}_{l}} \, , \\
    {\cal J}^{(OF)}_{l,l^\prime} & =\frac{(c-l) e^{-\beta l l^\prime \kappa_2} O_l}{c \hat{F}_{l}} \, , \\
    {\cal J}^{(OR)}_{l,l^\prime} & =\frac{l e^{-\beta (l-1) l^\prime \kappa_2} O_l}{c \hat{R}_{l-1}} \, , \\
    {\cal J}^{(RF)}_{l,l^\prime} & =\frac{(c-l) e^{-\beta (l+1) l^\prime \kappa_2} R_l}{c \hat{F}_{l+1}} \, , \\
    {\cal J}^{(RR)}_{l,l^\prime} & =\frac{l e^{-\beta l l^\prime \kappa_2} R_l}{c \hat{R}_{l}} \, .
    \end{aligned}
\end{equation}
The auxiliary functions $\hat{E}_q,\hat{O}_q,\hat{F}_q,\hat{R}_q$ are defined in Eq.~\eqref{eq:aux}.
In terms of these matrices, the elements of the Jacobian computed at the homogeneous fixed point are given by
\begin{equation} \label{eq:Japp}
    \begin{aligned}
    {\cal J} = \left . \frac{\partial \varphi_l^{(i \to j_0)}}{\partial \psi_{l^\prime}^{(j_1 \to i)}} \right \vert_{\rm h.s.} & = {\cal J}^{(\varphi \psi)}_{l,l^\prime} - \varphi_l \sum_{\chi \in \{E,F,O,R\}} \sum_{m=0}^c {\cal J}^{(\chi \psi)}_{m,l^\prime}  \, ,
    \end{aligned}
\end{equation}
where $(\varphi,\psi,\chi) \in \{E,F,O,R\}$ and all cavity fields take values corresponding to the solution of the homogeneous Eqs.~\eqref{eq:cavity}. 
The eigenvalues of the Jacobian can be computed numerically, and from the eigenvalue with the largest modulus, $\lambda_\mathrm{max}$, the linear and spin-glass instabilities of the paramagnetic phase can be determined by solving $c |\lambda_{\rm max}| = 1$ and $c \lambda_{\rm max}^2 = 1$, respectively. 

\bibliography{SALR}

\end{document}